\documentclass[12pt]{article}
\pdfoutput=1
\usepackage{arxiv}

\usepackage[utf8]{inputenc} 
\usepackage[T1]{fontenc}    
\usepackage{url}            
\usepackage{booktabs}       
\usepackage{amsfonts}       
\usepackage{nicefrac}       
\usepackage{microtype}      
\usepackage{lipsum}
\usepackage{graphicx}
\graphicspath{ {./images/} }

\usepackage{amsmath}
\usepackage{graphicx,psfrag,epsf}
\usepackage{enumerate}
\usepackage[numbers]{natbib}
\usepackage{optidef}
\usepackage{graphicx}
\usepackage{amsfonts}
\usepackage{mathrsfs}
\usepackage{amsmath}
\usepackage{enumerate}
\usepackage[table]{xcolor}
\usepackage{float}
\usepackage{subfigure}
\usepackage{lineno}
\usepackage{adjustbox}
\usepackage{siunitx}
\usepackage{scalerel}
\usepackage{booktabs}
\usepackage{courier}
\usepackage{multirow}
\usepackage{caption}[2007/11/04]
\usepackage{longtable}
\usepackage{setspace}
\usepackage{threeparttable}

\DeclareCaptionFont{blah}{\small\fontseries{n}\fontfamily{phv}\selectfont}

\DeclareCaptionLabelSeparator*{twonl}{\\[0.5\baselineskip]}

\newcommand{\vect}[1]{\boldsymbol{#1}}

\newcommand\vfrac[2]{\ThisStyle{%
  \setbox0=\hbox{$\SavedStyle#1#2$}%
  \setbox2=\hbox{$\SavedStyle X$}%
  \ifdim\ht0>\ht2\setlength{\ht0}{\ht2}\fi%
  #1\mathord{\stretchto{\raisebox{2.3\LMpt}{$\SavedStyle/$}}{\ht0}}#2}}

\newcommand{\E}{\mbox{$\textrm{\textup{E}}$}}
\newcommand{\Var}{\mbox{$\textrm{\textup{Var}}$}}
\newcommand{\Cov}{\mbox{$\textrm{\textup{Cov}}$}}

\newcommand{\bu}{\mbox{$\boldsymbol{u}$}}

\begin{document}

\title{Measurement Errors in Semiparametric Generalized Regression Models}

\author{
 Mohammad W. Hattab \\ School of Medicine  \\ The Johns Hopkins University \\ 
  \hspace{.3cm} \and
 \bf{David Ruppert}\\ School of Operations Research \\ and Information Engineering\\
 Department of Statistics  and Data Science \\
   Cornell University
}
\date{}
\maketitle

\begin{abstract}
 
 Regression models that ignore measurement error in predictors may produce highly biased estimates leading to erroneous inferences. It is well known that it is extremely difficult to take measurement error into account in Gaussian nonparametric regression. This problem becomes even more difficult when considering other families such as binary,  Poisson and negative-binomial regression.  We present a novel method aiming to correct for measurement error when estimating regression functions.  Our approach is sufficiently flexible to cover virtually all distributions and link functions regularly considered in generalized linear models. This approach  depends on approximating the first and the second moment of the response after integrating out the true unobserved predictors in any semiparametric generalized regression model.  By the latter is meant a model with both linear and nonparametric effects that are connected to the mean response by a link function and  a response distribution in an exponential family or quasilikelihood model.  Unlike previous methods, the method we now propose is not restricted to truncated splines and can utilize various basis functions. Through extensive simulation studies, we study the performance of our method under many scenarios.

\end{abstract}

\noindent%

\section{Introduction}
\label{sec:intro}
Consider the generalized semiparametric model
\begin{align}
E(y_i|\gamma, s) &= \mu\left(\vect v_i^\top  \vect \gamma + f(x_i)\right),\label{eq:01}
\end{align}
where $y_i$ is the outcome or response variable, $\mu$ is a monotonic and differentiable link function, $\vect v_i$ is a $p$-dimensional vector of covariates that enter the model linearly, $\vect \gamma$ is a $p$-dimensional parameter vector, $x_i$ is a scalar covariate, and $f$ is a smooth function.  This paper studies the challenging case where $x_i$ is measured with error.  Typically, it assumed that $y_i$ has a distribution in an exponential family which determines $\mu$.  Alternatively, a quasilikelihood can be defined by choosing $\mu$ and a variance function $V$ as in (6) below.  Extensions of model (1) are discussed in Section~5.

We will model $f$ as spline with a roughness penalty.   The most common penalty is 
\begin{align} 
\lambda \int \{f^{(2)}(x)\}^2  dx \label{eq:penalty2nd}
\end{align}
 where 
 $f^{(2)}$ is the second derivative of $f$ and
 $\lambda$ determines the strength of the penalty.  
 For concreteness, we will assume penalty (2), but it is straightforward to use a more general form of penalty.
 Notice that linear functions are not penalized under penalty~(2).

Assume that the spline basis is $(\vect x(x),\vect z(x))$ where $\vect x(x)= [1 \ x]$ consists of a basis of linear functions which are not penalized and $\vect z(x)$ consists of the $k$ basis functions that are penalized.  Then $f(x) = \vect x(x)^\top\vect \beta + \vect z(x)^\top \vect u $ for some parameter vectors $\vect \beta$ and $\vect u$.
Penalty (2) can be written as
\begin{align}
\lambda \int \{f^{(2)}(x)\}^2  dx &=
\lambda\vect u^\top \vect S \vect u,\label{eq:penaltyWithS}
\end{align}
where $\vect S $ is the  $k \times k$ penalty matrix whose $i,j$th entry is $\int z_i^{(2)} (s) z_j^{(2)} (s)ds$.  
Because the penalty is expressed using only the penalized component $\vect z$, $\vect S$ is positive definite, not merely positive semidefinite.
To estimate $\vect \beta$ and $\vect u$ we minimize the objective function defined as minus twice the log-likelihood
(or log-quasilikelihood) plus the penalty~(3).

As discussed in Ruppert, Wand, and Carroll (2003) and Wood (2017), the objective function is also minus twice the log-likelihood for a mixed model where $\vect u$ is given the distribution
\begin{align}
\vect{u} &\sim \text{N}(\vect 0,\phi   \vect S^{-1}/\lambda  ),
\end{align}
since the contribution to minus twice the log-likelihood due to $\vect u$ is (3).
Here $\phi$ is a scale parameter. For example, it represents the error variance for the Gaussian family, equals $1$ for binomial and Poisson families, and it allows for overdispersion in quasifamilies. To simplify notation, subsume $\vect v_i$ in (1) into $\vect x_i := \vect x(x_i)$ and $\vect\gamma$ into $\vect \beta$.  Define $\vect z_i = z(x_i)$.
Then our model is
\begin{eqnarray}
\E( y_i|\vect x_i, \vect z_i,\vect u) &=&  \mu(\vect{x}_i^\top\vect \beta + \vect {z}_i^\top \bu),  \label{eq:meanFn} \\
\Var(y_i|\vect x_i, \vect z_i,\bu)  &=&  V(\vect{x}_i^\top\vect \beta + \vect {z}_i^\top \bu, \phi, \theta), \quad \text{and} \label{eq:varFn} \\
\vect{u} &\sim& \text{N}(\vect 0,\phi  \vect S^{-1}/\lambda )  \label{eq:uDist}
\end{eqnarray}
for $i=1, \hdots ,n$, where the distribution of $y|\vect x,\vect z, \vect u$ corresponds to an exponential family or a quasi-likelihood, $\mu(\cdot)$ is the mean function in (1), $V(\cdot)$ is a non-negative differentiable function, $\vect \beta$ is a vector of fixed-effect regression parameters, $\vect {u}$ is a vector of unobservable random effects,  and $\theta$ is a variance parameter. The relationship between $\mu$ and $V$ is dictated by the assumed distribution. For Poisson models, $\mu=V$.  The parameters  $\phi$ and $\theta$ allow  quasilikelihood families and cases that aim to model more complex variance structures than the ones implied by regular members of the exponential family.  

As mentioned above, the unpenalized and penalized parts of the basis functions are represented by $\vect x$ and $\vect z$, respectively. For instance, a  truncated linear spline model implies that $\displaystyle \vect x_i^\top = [ 1 \quad x_i ] $ and $\displaystyle \vect z_i^\top = [(x_i - k_j)_+]^{j=1,\dots, k}$ where $\{k_1, \dots, k_k \} $ is a fixed set of knots and $x_+ = \max(x,0)$.
Truncated lines are not stable numerically, but the penalty ameliorates this problem.  More importantly, our approach allows us to use other bases.

We will be considering the situation where $x_i$ is not observed, but rather we observe $w_i$ which is $x_i$ plus measurement error.  This case violates
an important assumption in standard parametric and semiparametric regression models that predictors are measured without errors. Highly biased estimates may result if one fits a naive model that do not take measurement errors into account when they exist. Accordingly, severely misleading inferences will be produced. This is true regardless of the sample size. Carroll~(1989),  Cook  and  Stefanski  (1994),  Spiegelman, Rosner and Logan (2000) discuss measurement error in the parametric setting. Fuller~(1987) and Carroll et al.\ (2006) are devoted to the topic of measurement errors. 

As indicated by Berry, Carroll, and Ruppert~(2002), correcting for measurement error in nonparametric regression is an extremely difficult problem. Notable references that discuss the Gaussian case include  Fan and Truong~(1993), Carroll, Maca, and Ruppert~(1999), Berry et al.~(2002), Staudenmayer and Ruppert~(2004), and Sarkar, Mallick, and Carroll~(2014). Specifically, Berry et al.~(2002) developed an attractive Bayesian approach using linear truncated splines and assuming the prior distribution of the unobserved predictor is normal. The observed data likelihood function in Berry et al.~(2002) cannot be computed analytically as noticed by Ganguli, Staudenmayer and Wand (2005). Instead of seeking the distribution of the observed data, Hattab and Ruppert~(2021) found the exact mean and covariance of this distribution and accordingly developed an iterative heterosedastic mixed model method to estimate the regression parameters. This approach outperformed Staudenmayer and Ruppert~(2004) method and a set of other local polynomial estimators and is highly competitive with the Bayesian approach.

Nonparametric regression for other exponential family members has received far much less attention than the Gaussian case. This is possibly because the problem becomes tremendously more difficult. It is not clear how to generate posterior samples of the unobserved predictor for the Bayesian approach and it seems very tedious to find the exact moments of the observed data as done in Hattab and Ruppert~(2021).  

Let $w$ be the observed predictor. A classical assumption in measurement error models is that 
\begin{equation}
    w_i|x_i  \sim \text{N}(x_i,\sigma_w^2).
\end{equation}
Clearly, $\sigma_w^2=0$ implies that $x$ is measured without error. Usually, $\sigma_w^2$ is given or estimated using an external data-set or by the pooled sampled variance if replicates are available.  

There are two approaches to modeling the unobserved true predictors, $x_1,\dots,x_n$ (Carroll et al., 2006).  In so-called functional models, the predictors are assumed to be fixed (non-random) constants whereas in structural models, they are assumed to have been sampled from a distribution.
Following  Berry et al.~(2002), and Ganguli et al.~(2005), we will use the structural assumption that $x_i \sim \text{N}(\mu_x,\sigma^2_x)$; for $i=1,\hdots , n$ and some unknown $\mu_x$ and $\sigma^2_x$.  Hence, 
 \begin{equation}
  x_i| w_i \sim \text{N}\left(\frac{\sigma_x^2  w_i + \mu_x \sigma_w^2 }{\sigma_x^2 + \sigma_w^2} , \frac{\sigma_x^2\sigma_w^2}{\sigma_x^2 + \sigma_w^2}  \right) \quad \text{and} \quad  w_i\sim\text{N}\bigl(\mu_x, (\sigma_x^2 + \sigma_w^2) \bigr). \label{eq:xGivenw}
\end{equation} 
Later, we will relax the assumption of normality of $x$. 

Using linear truncated penalized splines and assuming $x$ having a normal distribution, Hattab and Ruppert~(2021) found the exact analytical form of $\E(y|w,\vect u)$ and $\Cov(y|w,\vect u)$ and developed a  heterosedastic mixed model method to estimate the regression parameters when the distribution in (1) is normal with identity link. However, the exact computations seem infeasible if the response distribution is not normal, or the link function is not identity, or  when using other basis functions such as B-splines, or when $x$ does not have a normal distribution. In this article, we  address all these limitations and extend the method of Hattab and Ruppert~(2021) to the semiparametric generalized regression models defined in (1).

In Section~2, rather than deriving exact expressions, we present a simulation-based method to estimate the mean and the covariance matrix of $y|w,\vect u$. 
Accordingly, a heterosedastic semiparametric model is devised. 
This method for correcting for covariate measurement error applies to any family or link function that falls under the model in (1)  including multinomial, quasilikelihood and Tweedie families with no restrictions on the distribution of $x$. 
Moreover, various basis representations such as cubic regression splines and thin plate regression splines can be utilized. Section~3  studies the performance of this new method through extensive Monte-Carlo simulations across many different scenarios. We restrict our study to non-Gaussian families since the Gaussian case has been extensively discussed in Hattab and Ruppert~(2021). For the binomial family, we show that our estimator is superior to a Bayesian probit estimator. An application to wage-union data is given in Section~4. 
Section~5 discusses extensions of model (1), or, equivalently, (5)--(7).
Conclusions are given in Section~6. We have found that our method corrects for bias across different distributions and regression functions, even for small samples.

\section{Methodology}
\label{sec:meth}
As indicated by Ruppert et al.~(2003) and Ganguli et al.~(2005), it is not possible to find the exact distribution of $y_i|w_i$ even when the distribution in (1) is normal which is the simplest case. Instead, Hattab and Ruppert~(2021) derived the exact expressions of $\E(y_i|w_i,\vect u)$ and $\Var(y_i|w_i,\vect u)$ when linear truncated splines are utilized. These quantities are very difficult to compute when considering other exponential families, link functions or basis representations. We will resort to simulations to approximate the first and the second moments. 
We will use the notation  $\vect Y=[y_i]$ and $\vect w=[w_i]$.
An immediate advantage of not integrating out the random effect $\vect u$ is being able to work with $\Cov(\vect Y|\vect w,\vect u)$, which is a diagonal matrix, instead of $\Cov(\vect Y|\vect w)$ which is a non-sparse matrix and difficult to work with especially when $n$ is large. This simple but very simplifying result applies to all models considered in~(5)--(7).   

First, the conditional mean of $y|w,\vect u$ is given by
\begin{eqnarray}
    \E(y_i| w_i, \vect u)  &=& \E\bigl(\E( y_i| w_i, \vect u, x_i)\bigr) 
    \nonumber\\&=& \E\left( \mu(\vect{x}_i^\top\vect \beta + \vect {z}_i^\top \bu)| w_i, \vect u\right).  \label{eq:EE}
\end{eqnarray}
To save ink, assume $\vect r_{i}^\top = [\vect x_{i}^\top,\vect {z}_{i}^\top]$ and $\vect b^\top = [\vect \beta^\top, \vect u^\top]$.
The quantity in~(10) cannot be found analytically but it is reasonably approximated by
\begin{eqnarray}
\E(y_i| w_i, \vect b) =  \E(\mu(\vect{r}_{i}^\top\vect b)| w_i, \vect b)  &\approx& g^{-1}\sum_{s=1}^{g} \mu(\vect{r}_{is}^\top\vect b), \label{eq:YgivenWandB}
\end{eqnarray}
where $x_{is}$ is sampled from $x_i|w_i$ and $\vect{r}_{is}^\top = [\vect x_{is}^\top,\vect {z}_{is}^\top]$. This quantity is further approximated using Taylor series expansion around an initial value $\vect b_0$
\begin{eqnarray}
\E(y_i| w_i, \vect b)  &\approx& g^{-1}\sum_{s=1}^{g} \mu(\vect{r}_{is}^\top\vect b_0) + g^{-1}\sum_{s=1}^{g} \frac{\partial \mu(\vect{r}_{is}^\top\vect b)}{\partial \vect b} \Bigr|_{\substack{\vect b=\vect b_0}} \vect r_{is}^\top (\vect b - \vect b_0)
\nonumber\\&=& g^{-1}\sum_{s=1}^{g}\Bigl(  \mu(\vect{r}_{is}^\top\vect b_0) - \frac{\partial \mu(\vect{r}_{is}^\top\vect b)}{\partial \vect b} \Bigr|_{\substack{\vect b=\vect b_0}} \vect r_{is}^\top  \vect b_0\Bigr) +g^{-1}\sum_{s=1}^{g} \frac{\partial \mu(\vect{r}_{is}^\top\vect b)}{\partial \vect b} \Bigr|_{\substack{\vect b=\vect b_0}} \vect r_{is}^\top  \vect b.  \label{eq:DR6}
\end{eqnarray}
The first term in the right hand-side is an offset term and denoted by $\vect O = [O_i]$. The model matrix, which we will denote by $\vect M$, is the matrix that pre-multiplies  $
\vect b$ in (12) and is partitioned as $\displaystyle \vect M=[\vect M_\beta, \vect M_u]$ correponding to the partition  $\vect b= [\vect \beta^\top \ \vect u^\top]^\top$. The matrix $\vect M_\beta$ corresponds to $\vect \beta$ and is not penalized whereas the penalized part is given by $\vect M_u$. For example, in linear penalized splines, $\vect M_\beta$ is $n\times 2$ matrix and $\vect M_u$ is $n\times k$ matrix. 

Next, the approximation of the conditional variance of $\vect y|w,\vect b$ is given by
\begin{eqnarray}
    \Var(y_i| w_i, \vect b)  &=& \E\bigl(\Var( y_i| w_i, \vect b, x_i)\bigr) + \Var\bigl(\E(y_i|w_i,\vect b, x_i)\bigr) 
    \nonumber\\&=& \E( V( \vect {r}_i^\top \vect b,\phi, \theta)| w_i, \vect b) + \Var\bigl(\mu(\vect {r}_i^\top \vect b)\bigr)
     \nonumber\\&\approx& g^{-1} \sum_{s=1}^{g}  V( \vect {r}_{is}^\top \vect b,\phi, \theta) + \widehat{\Var}\bigl(\mu(\vect {r}_{is}^\top \vect b)\bigr).\label{eq:DR7}
\end{eqnarray}
 Under the normal case, the first term is simply the error variance. For the Poisson and quasi-Poisson with log-link, it is $\displaystyle g^{-1} \sum_{s=1}^{g}  \exp( \vect {r}_{is}^\top \vect b)$ and $\displaystyle \phi g^{-1} \sum_{s=1}^{g}  \exp( \vect {r}_{is}^\top \vect b)$, respectively, and it is \break $\displaystyle \theta g^{-1} \sum_{s=1}^{g}  \exp(2 \vect {r}_{is}^\top \vect b) + g^{-1} \sum_{s=1}^{g}  \exp( \vect {r}_{is}^\top \vect b)$ under the negative binomial family. For the Bernoulli family, unlike other families, the exact distribution of $y|w, \vect b$ is known. Hence, the  variance can be approximated by  $\displaystyle g^{-1}\sum_{s=1}^{g} \mu(\vect{r}_{is}^\top\vect b) \times  \left(1- g^{-1}\sum_{s=1}^{g} \mu(\vect{r}_{is}^\top\vect b)\right) $. The second  term in (13) is the sample variance computed from the samples $\mu(\vect {r}_{is}^\top \vect b)$. Regardless of the assumed family, this term implies heteroscedasticity as the variance varies with $i$. 

As mentioned earlier, $\Cov(\vect Y|\vect w, \vect u) $ is a a diagonal matrix. To see this, note that  this matrix is the sum of two components, $\E\bigl(\Cov(\vect Y|\vect x,\vect w, \vect u)|\vect w, \vect u\bigr)$ and $\Cov\bigl(E(\vect Y|\vect x,\vect w, \vect u)|\vect w, \vect u \bigr)$. The response $\vect Y$ is independent of $\vect w$ given $\vect x$ and $\Cov(\vect Y|\vect x, \vect u)$ is a diagonal matrix according to model (5)--(7), since $\vect Y|\vect x, \vect u$ is a regular GLM (generalized linear model). The $j$-th element of the random vector $E(\vect Y|\vect x,\vect u)$ depends on $x_j$ alone and since $x_i$'s are independent given $w_i$'s the covariance of $E(\vect Y|\vect x, \vect u)$ is zero everywhere except on the diagonal.        

Typically, $\sigma_w^2$ is given. It may be estimated using an external data-set or by the pooled sampled variance if replicates are available. Regarding the parameters $\mu_x$ and $\sigma_x^2$, if $x$ follows a normal distribution, the method of moments will be used. Specifically, $\mu_x$ is estimated by the sample mean of $\vect w$ and $\sigma_x^2$ is estimated by the sample variance of $\vect w$ minus  $\sigma_w^2$. Once these estimators are obtained, they are plugged in Equation~(9) and remain fixed afterwards.

The distribution of $x_i|w_i$ is given in (9) if $x$ has a normal distribution. According to (9), this assumption can be investigated using $w_i$'s. In the next section, we show that our estimator still performs well even if this assumption is severely violated. However, if one is not willing to assume that $x$ has a normal distribution or any other specific distribution, then we suggest proceeding as follows to simulate from $x|w$ distribution. First, the density of $x$ is estimated using one of the density deconvolution methods such as the Fourier based kernel density estimator (Stefanski and Carroll~1990; Diggle and Hall~1993). The sampling weights to sample from the distribution of $x|w$ are given by multiplying the estimate of the density of $x$ by the density of $w|x$. Despite its simplicity, we found that Stefanski and Carroll's~(1990) estimator where its bandwidth is computed by the plug-in method of Delaigle and Gijbels~(2002) provides more stable predictions than when using the penalized contrast method (Comte, Rozenholc, and Taupin~ 2006), the Bayesian estimator of Sarkar et al.~(2014), and the quadratic programming estimator (Yang et al.~2020).

Finally, the Equations in (12) and (13) suggest formulating the problem in the form of a linear heterosedastic mixed model. Specifically,
\begin{equation}
\vect{Y} - \vect {O} =  \vect M_\beta \vect \beta + \vect M_u \vect u + \vect {\epsilon},   \quad  \vect{u} \sim \text{N}(\vect 0,\phi  \vect S^{-1}/\lambda ), \label{eq:DR8}
\end{equation} where $\vect {\epsilon}$ is a random vector that has unknown distribution with mean $\vect 0$ and diagonal covariance matrix with $\displaystyle \Var(y_i| w_i, \vect u)$ as approximated in (13) being on the diagonal. Notice that $\E(\vect {\epsilon}| \vect u ) = \vect 0 $ and the covariance of $\vect {\epsilon}$ depends on $\vect \beta$ and $\vect u$ in a non-linear fashion.   

Based on the previous discussion and the formulation in~(14), we propose the following iterative algorithm:

\begin{enumerate}
\item For each $i=1,\ldots, n$, generate $g$ random values from the distribution of $x_i|w_i$ as specified in Equation~(9) if $x_i$ has a normal distribution  or nonparametrically  based on a deconvolution estimator.

\item Fit a naive  semiparametric GLM with $w$ the observed predictor replacing the true predictor $x$ in (1). Utilizing this fit, extract the basis function representation of the simulated values in step 1, $\vect r_{is}$,  and the associated penalty matrix $\vect S$.

\item From the fitted model, extract the regression parameters estimate, $\vect b_0$. 
\item Using $\vect b_0$ from the previous step, form the matrices $\vect O$, $\vect M_\beta$, and $\vect M_u$. Plug-in $\vect b_0$ in (13): find $\displaystyle V(\vect r_{is}^\top \vect b_0,\phi, \theta)$ and compute an estimate of the  variance of $\mu(\vect {r}_{is}^\top \vect b_0)$.
\item Fit the linear heteroscedastic mixed model in (14) with variance weights being estimated from the previous step according to (13) at $\vect b = \vect b_0$. Note that the second term in (13) is considered known but varies with $i$. Depending on the assumed family, the first term in (13)  is considered completely known and varies with $i$ as in the Poisson and binomial families, unknown and does not change across $i$ as in the normal case, or partially unknown and varies with $i$ as in the gamma, negative binomial,  quasi-Poisson and quasi-binomials families.  
\item Repeat steps 3--5 until stabilization. 
\item Using the estimates from the previous step, obtain the fitted curve over a grid of values $\vect d = (d_1,\ldots ,d_t)$
 \[  \mu(\vect{X}_d^\top\hat{\vect \beta} + \vect {Z}_d^\top \hat{\bu}) \] 
 where $\vect{X}_d$ and $\vect{Z}_d$ are the basis representation of $\vect d$ and $\mu$ is applied component-wise.
\end{enumerate}

 This approach will be called by the observed semiparamteric measurement error estimator (OSMEE) at it revolves around the observed values of the predictor after integrating out the true predictor $x$.

 A major consideration is how to estimate the smoothing parameter $\lambda$. For this purpose, there are two strategies. The first one is to fix $\lambda$ in advance through all iterations and compute the solution path accordingly. This process is repeated over a grid values of $\lambda$ and an optimal value is chosen via $k$-fold cross validation. This is evidently can be time consuming. A much faster alternative which we will adopt is to a apply smoothness selection criterion to the working model in step (5) at each iteration. The Gaussian version of the generalized cross validation (GCV) or the restricted maximum likelihood (REML) can be used for this purpose. This strategy is similar to the penalized quasi likelihood (PQL) (Breslow and Clayton, 1993, Ruppert et al.\ 2003, and Wood 2017) method in generalized linear mixed models. The GCV can be easily justified (see Wood~2017). However, we found that it may produce highly unstable results. On the other hand, it is not clear how REML can be justified since the distribution $y|w,\vect u$ is not Gaussian. Nonetheless, REML consistently shows a very good performance across distributions and sample sizes as it will be seen in the next section.

 We noticed that the estimates at the convergence might suffer from over-fitting, i.e $\lambda$ is too small. To refine the estimates, a modified version of the GCV formula that corresponds to the model in (5)--(7) is computed for each iteration. Specifically, this quasi-GCV score is computed as 
  \begin{equation}
 \text{QGCV}_j = n D(\hat{\vect b}_{j})/(n-\eta_{j})^2\label{eq:DR9}  \end{equation}
 where $D(\cdot)$ is deviance of the distribution of $y|x,\vect u$, and $\hat{\vect b}_{j}$ and $\eta_{j}$ are the parameter estimate and the effective degrees of freedom (EDF), respectively, for the linear mixed model at the $j$th iteration. The fitted values in $D(\cdot)$ are computed by plugging-in $\hat{\vect b}_{j}$ in (11), i.e. $g^{-1}\sum_{s=1}^{g} \mu(\vect{r}_{is}^\top\hat{\vect b}_{j})$. Also, the EDF, $\eta_j$, is the trace of the influence matrix that is given by 
 \begin{equation}   
(\vect C_j^\top \vect W_j \vect C_j + \hat \lambda_j \vect \Psi)^{-1} \vect C_j^\top \vect W_j \vect C_j \end{equation}
 where $\vect C_j = [\vect M_\beta : \vect M_u ]_j$, $\vect {W_j}$ is a diagonal matrix carrying the inverse of the estimated variance weights estimated, $\hat \lambda_j$ is the REML estimate produced by the working model in step 5, and $\vect \Psi$ is a block diagonal matrix containing $\vect S$ on the lower block and zero otherwise. The fit that corresponds to the iteration with the lowest QGCV score is selected and the corresponding parameter estimate is used to obtain the fitted curve in step 7. Details regarding REML, GCV, PQL, and EDF can be found in Wood~(2017).

 In terms of prediction accuracy, in the next section we will evaluate the differences between GCV and REML, the use of deconvolution estimator to estimate the density of $x|w$ versus assuming normality, varying the basis dimension, and using different basis representations. We will also compare our estimator with the Bayesian probit estimator developed by Berry et al.~(2002).

\section{Simulations}
\label{sec:sim}
In this section, we will evaluate the performance of our method through simulations under a variety of cases, sample sizes, and response distributions. Considering Poisson regression, subsections 3.1--3.3 study smoothing parameter selection criteria (REML vs.\ GCV), estimation of the density $x|w$, and basis functions and dimension. For binary data, subsection 3.4 contrasts the prediction accuracy of the method against the Bayesian estimator. Subsection 3.5 assesses the effectiveness of the method when the response variable has a  Gamma or negative-binomial distribution.   

Unless stated otherwise, the default smoothness selection criterion is REML with QGCV in (15) applied and the default basis function is thin plate regression splines (TP) with basis dimension set at $40$. The definition of this basis function and other basis functions discussed later and the penalty matrices associated with them are given in Wood~(2017).

We assume throughout that the measurement error variance $\sigma_w^2$ is known and needs not to be estimated. The mean squared error (MSE) is evaluated over a grid of $101$ points covering most of the range of true predictor $x$. The sample size varies from $2^7$ to $2^{11}$ with $300$ simulated data-sets generated at each sample size. Finally, we set $g=3000$ in step (1) of the algorithm.

\subsection{Smoothing Selection Criterion: REML vs. GCV}\label{subsec:REML_GCV}

\begin{figure}[!t]
\centering
\makebox{\includegraphics[trim=1 170 1 200,clip,width=1.00\textwidth]{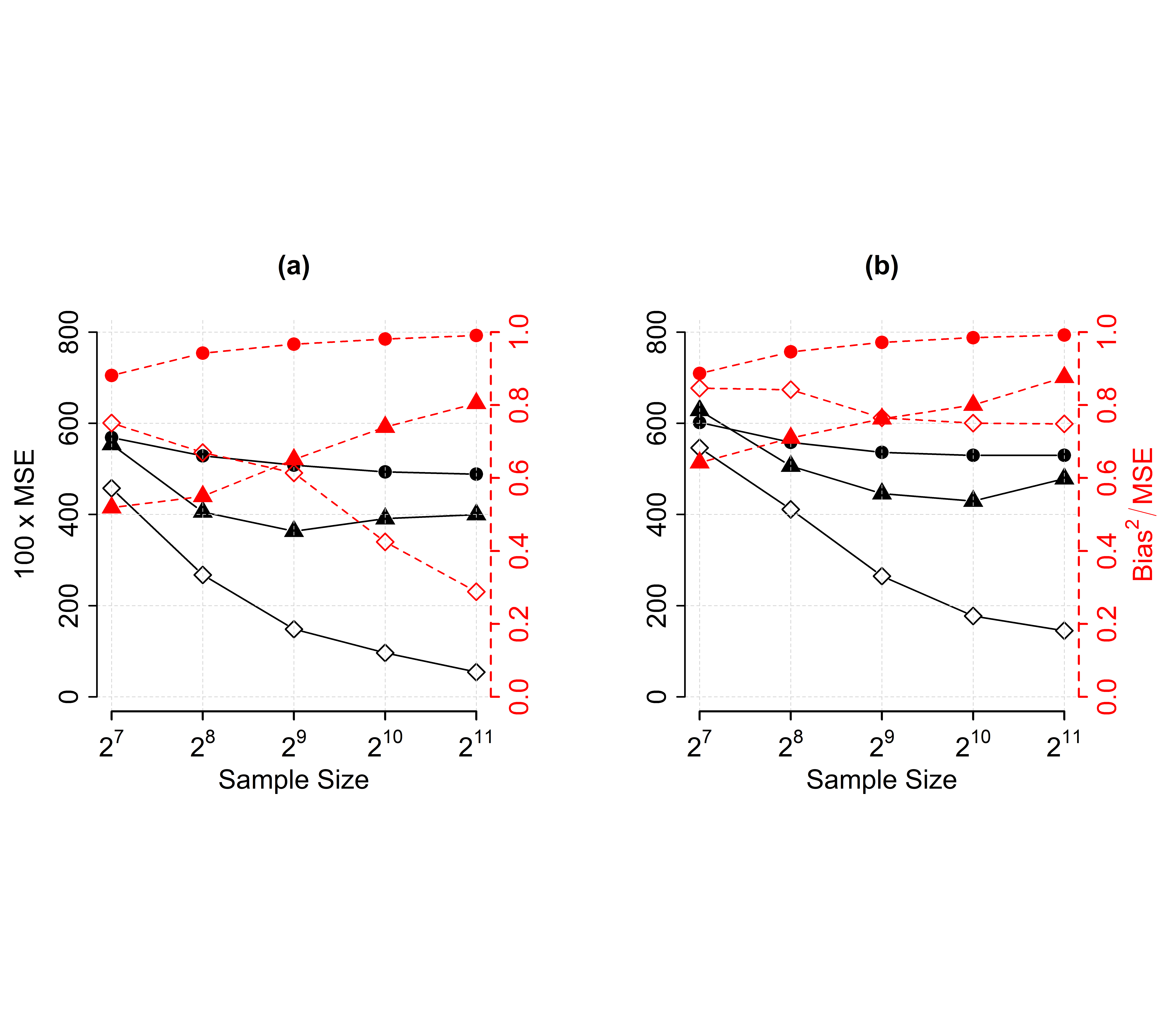}}
\caption{ MSE (black) for semiparametric Poisson regression computed for the naive approach (solid circles), OSMEE-REML (open diamonds), and OSMEE-GCV (solid triangles). The mean squared bias/MSE is shown in red. The true regression function is $\displaystyle \exp(2 \sin(4\pi x))$. (a) $x$ has a Gaussian distribution, (b) $x$ has a skew normal distribution with $\alpha=6$. }
\end{figure}

Consider Poisson regression with log-link. The true  function on the $\log$ scale is $\displaystyle 2 \sin(4\pi x)$. The mean and standard deviation of $x$ distribution are $0.5$ and $0.25$, respectively. The measurement error variance  is $0.141^2$. The MSE is assessed over $101$ equally spaced points between $0.1$ and $0.9$.
 
Figure~1 shows the average MSE along with the percentage of the contribution of $\text{Bias}^2$ to the MSE for the OSMEE when adopting REML or GCV in the fitting process. Also, a TP (thin plate) naive model that does not take measurement error into account is shown. This is basically the result of step 2 in the algorithm. Panel~(a) draws simulations of $x$ from Gaussian distribution with mean $0.5$ and standard deviation $0.25$ whereas Panel~(b) draws from skew-normal (Azzalini ~1985 and Azzalini~2013) with the same mean and standard deviation but with shape parameter $\alpha=6$ which corresponds to a very long tailed distribution.  In other words, Panel~(a) corresponds to the case where the distribution of $x$ is correctly specified while it is incorrectly specified in Panel~(b) since step 1 of the algorithm assumes $x|w$ has a normal distribution. We will relax this assumption shortly.   

It is evident that REML largely outperforms GCV at all sample sizes. The former has a sturdy performance as it vastly improves as $n$ increases even when the distribution of $x$ is incorrectly specified. On the other hand, there is only a slight improvement for GCV beyond $n=2^{8}$. GCV has produced very unstable results for few samples when $n=2^{7}$ and $n=2^{8}$; a situation we did not encounter when using REML. Those samples were excluded from the analysis. It seems that GCV is prone to over-fitting which leads to poor predictions. For this reason we prefer REML.  

Notice that in Panel~(a) how the bias has dramatically decreased for REML while it overwhelmed the naive fit. In Panel~(b), as expected, distribution misspecification has introduced additional bias to the estimator as the approximation of $E(\vect Y|\vect w, \vect u)$ becomes less accurate and therefore the MSE has increased. Despite that, our method is still able to correct for the measurement error bias, showed a robust performance  against distribution misspecification, and it remained far superior than the naive approach for all sample sizes.

\subsection{$x|w$ Density} \label{subsec:density}

\begin{figure}[!t]
\centering
\makebox{\includegraphics[trim=1 170 1 200,clip,width=1.00\textwidth]{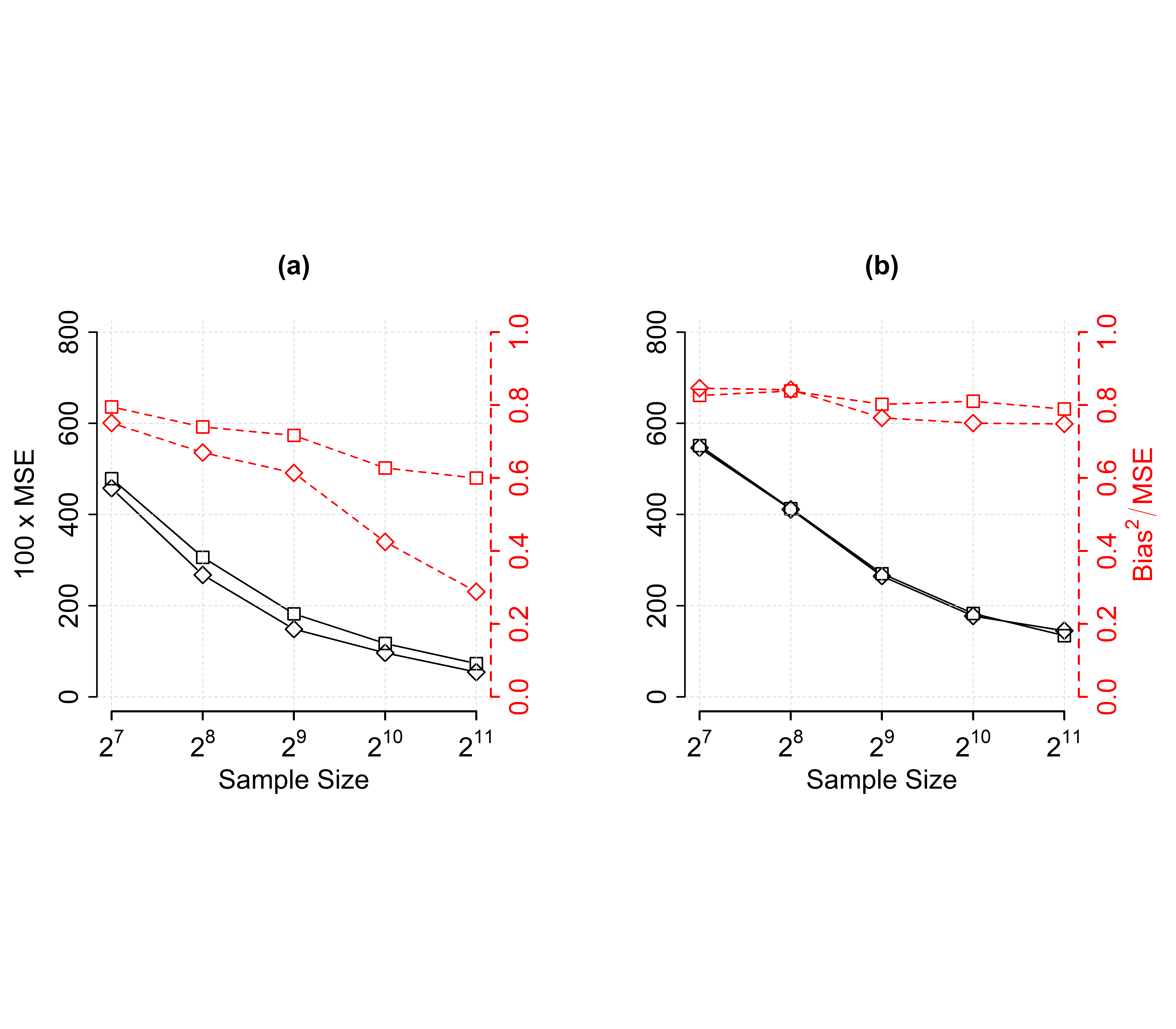}}
\caption{ MSE (black) for semiparametric Poisson regression computed for the OSMEE with Gaussian sampling (open diamonds), and OSMEE with deconvolution sampling (open squares). The mean squared bias/MSE is shown in red. The true regression function is $\displaystyle \exp(2 \sin(4\pi x))$. (a) $x$ has a Gaussian distribution, (b) $x$ has a skew normal distribution with $\alpha=6$. }
\end{figure}

 Step 1 of the algorithm samples $x|w$ from normal distribution depending on the assumption that $x$ is normally distributed. It is noticed from Figure~1 that the OSMEE is insensitive to a large extent to the normality assumption even when it is severely violated. Instead, one can use a deconvolution method to sample from $x|w$ without specifying a distribution for $x$ as described in the previous section. Specifically, the distribution of $x$ is estimated via the deconvolution kernel estimator given by Stefanski and Carroll~(1990) where its bandwidth is computed by the plug-in method of Delaigle and Gijbels~(2002). The estimated density of $x$ is multiplied by the density of $w|x$ to produce sampling weights required to sample from $x|w$. Figure~2 recomputes the simulation study in Figure~1 when now the distribution of $x$ is not specified. 

In Panel~(a), when the true distribution of $x$ is normal, the method based on normal sampling has a slight advantage over the deconvolution sampling. The difference between the two sampling methods become negligible when the true distribution of $x$ is skew normal as in Panel~(b). This may suggest favoring normal sampling because of its simplicity and its ability to correct for measurement error bias even when the normality assumption is violated at least in this example. However, as we will see in later sections, the difference can be noticeable unlike the situation here. Generally, we recommend simulating from normal unless there is an indication that the distribution of $x$ is far from normal. 

\subsection{Smoothing Basis}\label{subsec:SmoothingBasis}

\begin{figure}[!t]
\centering
\makebox{\includegraphics[trim=1 170 1 200,clip,width=1.00\textwidth]{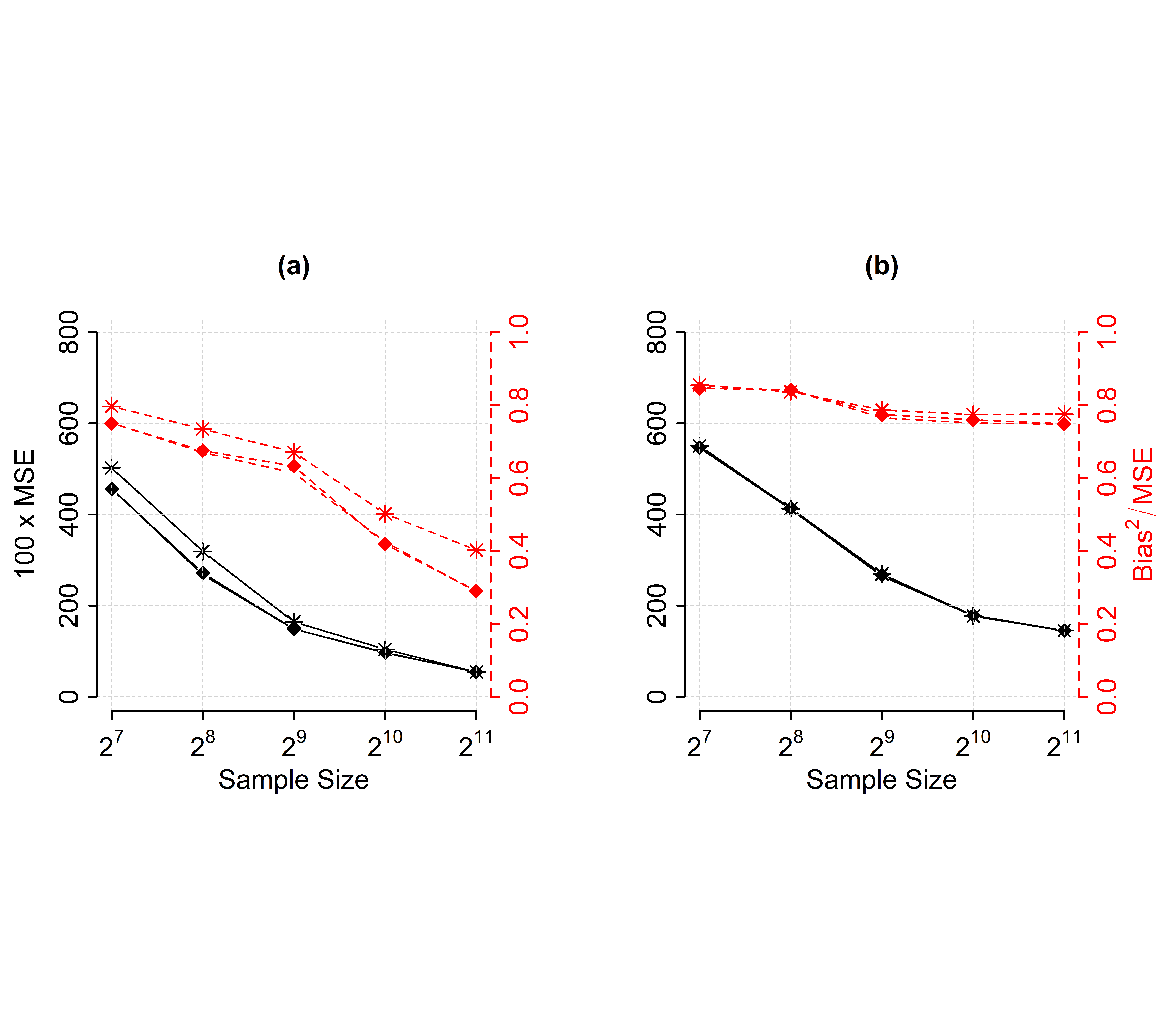}}
\caption{ MSE (black) for semiparametric Poisson regression computed for the OSMEE with basis dimension equals 10 (stars), 25 (closed diamonds), or 40 (open diamonds). The mean squared bias/MSE is shown in red. The true regression function is $\displaystyle \exp(2 \sin(4\pi x))$. (a) $x$ has a Gaussian distribution, (b) $x$ has a skew normal distribution with $\alpha=6$. }
\end{figure}

As mentioned earlier, the basis dimension for the TP was set at $40$ for the previous simulations. Figure~3 repeats the analysis in Figure~1 with now the basis dimension reduced to $10$ and $25$. To ease comparisons, Figure~3 also includes the previous results. There is a clear improvement in terms of the MSE and the bias when increasing the basis dimension to $25$ but there is a very little gain beyond that. In panel~(b) where the distribution of $x$ is misspecified, the differences between the three cases are virtually indistinguishable.

\begin{figure}[!t]
\centering
\makebox{\includegraphics[trim=1 170 1 200,clip,width=1.00\textwidth]{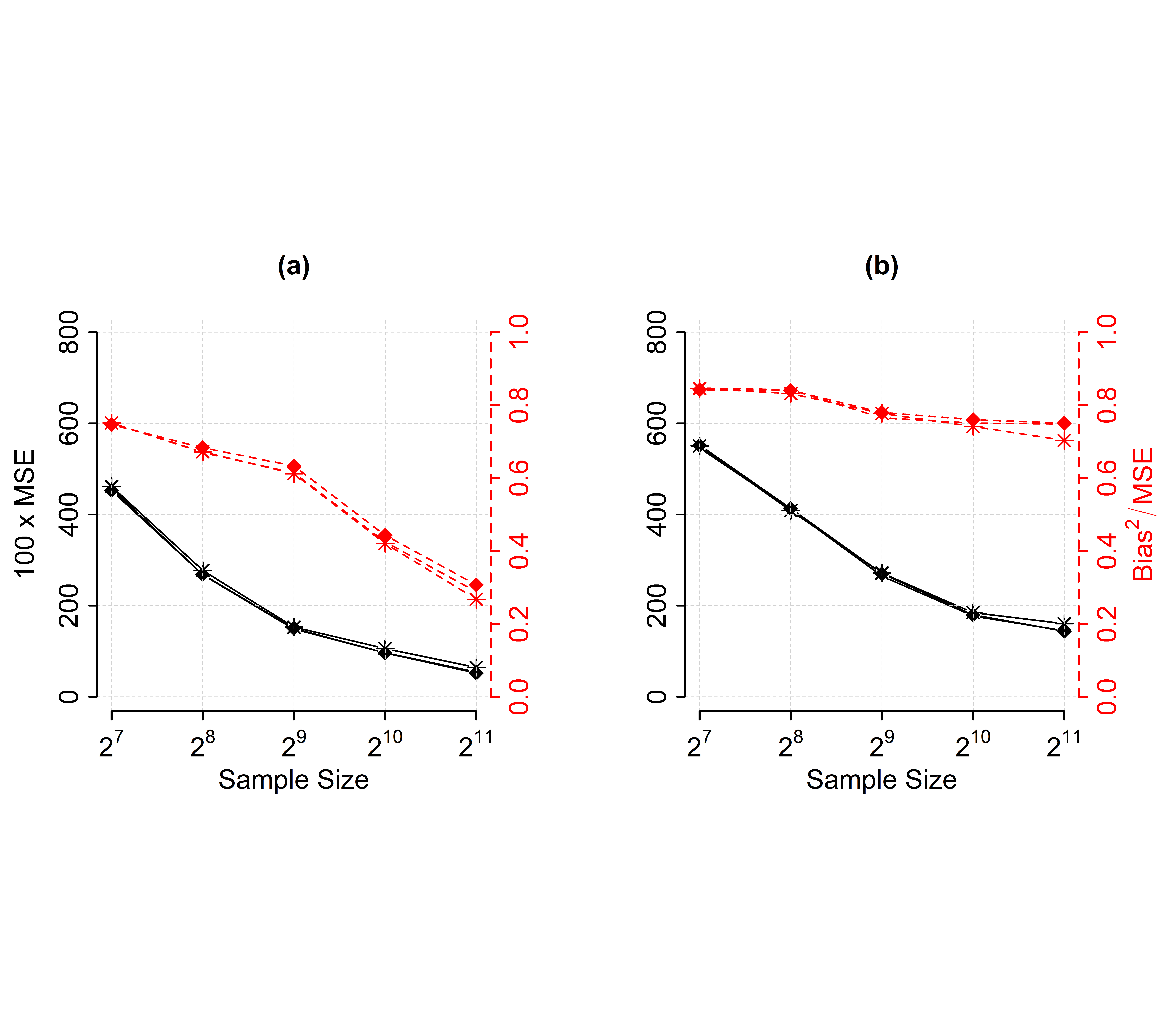}}
\caption{ MSE (black) for semiparametric Poisson regression computed for the  ME model with smoothing basis sets at  CR (stars), PS (closed diamonds), or TP (open diamonds). The mean squared bias/MSE is shown in red. The true regression function is $\displaystyle \exp(2 \sin(4\pi x))$. (a) $x$ has a Guassian distribution, (b) $x$ has a skew normal distribution with $\alpha=6$. }
\end{figure}

Next, next we vary the smoothing basis to include cubic regression splines (CR) and P-splines (PS). Figure~4 shows the results. It seems that performance of the method is almost the same when using any of the three basis functions with a very slight advantage of TP and PS over CR.

\subsection{Logistic Regression}\label{subsec:logistic}

In this section we will evaluate the efficacy of our method in adjusting the measurement error bias in nonparametric logistic regression. Four regression functions (listed below on the logit scale)  will be studied. The MSE is measured on the probability scale over a grid of $101$ points enclosed by $a$ and $b$ specified below covering most of the range of $x$. The sample size $n$ varies from $n=2^{7}$ to $n=2^{11}$ and $300$ data-sets are generated at each sample size. 

Similar to above, for each case, the simulation study is conducted twice; first assuming the true covariate $x$ has a normal distribution and then assuming $x$ has a skew normal distribution with the shape parameter is equal to $6$. Both distributions have the same mean and variance. Finally, we will compare the OSMEE to the Bayesian probit model developed by Berry et al.~(2002). It assumes that $x$ has a Gaussian prior distribution with non-informative priors imposed on $\mu_x$ and $\sigma_x$.

 The four cases considered as follows. (The parameters $\theta$ and $\gamma$ are the shape parameter for the negative binomial and gamma families discussed in the next section and they can be ignored for now.)

 \begin{enumerate}
    
    \item $a=0.1$, $b=0.9$, $\sigma_w^2 = 0.141^2$, $\mu_x = 0.5$, $\sigma_x^2 = 0.25^2$, $\theta=6$, $\gamma=2$, and the regression function 
    \begin{equation}
        m(x) = 2\sin(4 \pi x) \nonumber
    \end{equation}
    
    \item $a=-2$, $b=2$, $\sigma_w^2 = 0.8^2$, $\mu_x = 0$, $\sigma_x^2 = 1$, $\theta=3$, $\gamma=6$, and the regression function
     \begin{equation}
        m(x) = 2 \tanh(x) \nonumber
    \end{equation}
   
   \item $a=0.1$, $b=0.9$, $\sigma_w^2 = 0.11^2$, $\mu_x = 0.5$, $\sigma_x^2 = 0.25^2$, $\theta=1.5$, $\gamma=10$, and the regression function
     \begin{equation}
        m(x) = 100(x)_+^3 (1-x)_+^3 \nonumber
    \end{equation}
   
   \item $a=0.3$, $b=0.8$, $\sigma_w^2 = 0.0.075^2$, $\mu_x = 0.6$, $\sigma_x^2 = 0.12^2$, $\theta=5$, $\gamma=4$, and the regression function
     \begin{equation}
        m(x) =  2\exp(-60(x-0.6)^2) + 0.25/(0.1+x)\nonumber
    \end{equation} 
    
\end{enumerate}
 
Note that case~1 is the case presented previously for the Poisson regression. Case~4 is a slight modification of the ``bump function" in Ruppert et al.~(2003). Figure~5 shows the ratio of the MSE of the Bayesian approach to the MSE of the OSMEE. The OSMEE with Gaussian sampling is shown on the top panels and the bottom panels correspond to the OSMEE with deconvolution sampling. The distribution of the unobserved predictor $x$ is Gaussian in Panels~(a) and (c) and skew-normal in Panels~(b) and (d).  The OSMEE approach incorrectly specifies the distribution of $x$ in Panel~(b)  and the Bayesian approach incorrectly specifies it  Panels~(b) and (d).

\begin{figure}[!t]
\centering
\makebox{\includegraphics[width=1.00\textwidth]{f5}}
\caption{ The MSE ratio is computed as $\text{MSE}_{Ratio} = \text{MSE}_{Bayes}/\text{MSE}_{OSMEE}$  for cases 1--4 described in the text: case~1 (crosses), case~2 (diamonds), case~3 (inverted triangels), and case~4 (crossed squares). (a) $x$ has a Gaussian distribution and $\text{MSE}_{OSMEE}$ computed with Gaussian sampling, (b) $x$ has a skew normal distribution and  $\text{MSE}_{OSMEE}$ computed with Gaussian sampling, (c) $x$ has a Gaussian distribution and $\text{MSE}_{OSMEE}$ computed with deconvolution sampling, (d) $x$ has a skew normal distribution and  $\text{MSE}_{OSMEE}$ computed with deconvolution sampling. }
\end{figure}

The performance of the OSMEE has evidently dominated the Bayesian approach for all cases and sample sizes except when $n=2^{7}$ for cases 3 \& 4. The difference between the methods increases as the sample size increases and in some cases the difference is tremendous in favor of the OSMEE approach and in one case it is better by more than $8$ folds. When both approaches correctly specify the distribution of $x$, the average improvement is about $119.2\%$ and $82.5\%$ when they both misspecify it. These numbers change to $92.7\%$ and $92.8\%$ when the OSMEE is used with deconvolution sampling that does not specify a distribution for $x$. 

The difference between using the Gaussian sampling and the deconvolution sampling is more clear than what we have seen in Figure~2. Generally speaking, when the Gaussian assumption is violated, the deconvolution sampling reduces the bias induced in the estimator when using Gaussian sampling. Despite the serious lack of normality, the OSMEE with Gaussian sampling showed a strong performance. We will see more on this in the next section when discussing negative binomial regression and gamma regression.

\subsection{Negative-binomial and Gamma Regression}\label{subsec:negaBin}

In this section, we will re-run cases 1--4 assuming now the response variable has a negative-binomial distribution or gamma distribution with $\log$ link utilized. There is a wide variety of methods for adjusting for measurement error in Gaussian nonparametric regression. Many of those methods were discussed in Hattab and Ruppert~(2021). To the best of our knowledge, there are no other methods to adjust for measurement error when performing nonparametric negative-binomial regression or gamma regression or any nonparametric regression for that matter when the data are not Gaussian or binary. In fact, our method can be applied to any (parametric and non-parametric) glm family not only those discussed in this section including important cases such as inverse-Gaussian, beta, multinomial data (ordered and unordered responses), and Tweedie distributed data.

\begin{figure}[!t]
\centering
\makebox{\includegraphics[width=1.00\textwidth]{f6}}
\caption{ MSE (black) for semiparametric negative-binomial regression computed for the naive approach (solid circles), the OSMEE with Gaussian sampling (open diamonds), and OSMEE with deconvolution sampling (open squares). The mean squared bias/MSE is shown in red. The four cases are described in the text. The true predictor $x$ has a Gaussian distribution. }
\end{figure}

\begin{figure}[!t]
\centering
\makebox{\includegraphics[width=1.00\textwidth]{f7}}
\caption{ MSE (black) for semiparametric negative-binomial regression computed for the naive approach (solid circles), the OSMEE with Gaussian sampling (open diamonds), and OSMEE with deconvolution sampling (open squares). The mean squared bias/MSE is shown in red. The four cases are described in the text. The true predictor $x$ has a skew-normal distribution with $\alpha=6$. }
\end{figure}

\begin{figure}[!t]
\centering
\makebox{\includegraphics[width=1.00\textwidth]{f8}}
\caption{ MSE (black) for semiparametric gamma regression computed for the naive approach (solid circles), the OSMEE with Gaussian sampling (open diamonds), and OSMEE with deconvolution sampling (open squares). The mean squared bias/MSE is shown in red. The four cases are described in the text. The true predictor $x$ has a Gaussian distribution. }
\end{figure}

\begin{figure}[!t]
\centering
\makebox{\includegraphics[width=1.00\textwidth]{f9}}
\caption{ MSE (black) for semiparametric gamma regression computed for the naive approach (solid circles), the OSMEE with Gaussian sampling (open diamonds), and OSMEE with deconvolution sampling (open squares). The mean squared bias/MSE is shown in red. The four cases are described in the text. The true predictor $x$ has  a skew-normal distribution with $\alpha=6$. }
\end{figure}

The results are shown in Figures~6--9. For both families, our method seems to correct for bias and dominates the naive fit regardless of the distribution family and regardless if the distribution of $x$ is correctly specified or not for most sample sizes. Interestingly, in case--2 at $n=2^7$ the Gaussian sampling is inferior to the deconvolution sampling even when the distribution of $x$ is normal and is superior to the deconvolution sampling in case--3 for almost all sample sizes even when the distribution of $x$ is skew-normal. The Gaussian sampling shows a robust behavior as one is still able to account for measurement error even when the assumed model is wrong. Having said that, there is an overwhelming evidence found in case--2 for use of the deconvolution sampling.

For gamma regression, in cases 3 and 4, it seems there is a small loss in the method's performance when doubling the sample size from $n=2^{10}$ to $n=2^{11}$ for the Gaussian sampling when the distribution of $x$ is misspecified. Further investigation shows that  $\hat \lambda$ appears to be too small for theses cases resulting in under smoothing, a situation we did not encounter when using deconvolution sampling .

\section{The Wage-Union data: Sensitivity Analysis} \label{sec:WageUnion}

The wage data contain two variables, union membership (binary) and wages (continuous). Ruppert et al.~(2003) fits a logistic spline regression of union membership on wages using $20$-knots. We study the sensitivity of the fit to measurement error in wages. There is no information regarding the measurement error variance and a replicate of the data is not available. Our aim here is to assess the sensitivity of the fitted curve to measurement error in the predictor.  

Five measurement error variance are considered; $\sigma_w^2=0,1,4,9, \text{and}\ 16$. The reliability ratio is given by:
$$
\text{reliability ratio} = \Var(\text{wages})/\left(\Var(\text{wages})+\sigma_w^2\right)
$$
The sample variance of wages is $26.41$ and therefore the corresponding reliability ratio ranges from $1$ (no measurement error) to $0.62$. The regression functions that relate wages to union membership based on the OSMEE using Gaussian and deconvolution sampling  are shown in Figure~(10). The Bayesian fit of Berry et al.~(2002) is also included. 

The fitted curves appear to be sensitive to measurement error with degree of sensitivity varies from approach to approach. The OSMEE with Gaussian sampling varied the most and produced implausible fits when $\sigma_w^2=9$ and $16$. For $\sigma_w^2=16$, the probability of union membership is basically $0$  for individuals with wages greater than $15$ or less than $9$. The deconvolution sampling provides more stable fits. The Bayesian approach have suggested slight to moderate changes to the fitted curves as $\sigma_w^2$ increases. Note that how the location of the peak changed from about $\$12$ to $\$16$. Recall that, as demonstrated in the previous section, the Bayesian model was considerably inferior to the OSMEE and may miss important structure of the data. 

It is noted the distribution of wages is skewed to the right and this may explain the discrepancies between the Gaussian and the deconvolution sampling. As observed from the previous section, the Gaussian sampling was competitive with the deconvolution sampling in some occasions where the lack of normality is serious. In this case here, since the Gaussian sampling has produced unreasonable fits, the deconvolution sampling is to be preferred. 

After log-transformation, the variable wage appears fairly symmetric. The OSMEE with Gaussian sampling is applied on the transformed data. The measurement error variance is transformed as well maintaining the same reliability ratio as before. The results are shown in Panel~d of Figure~(10). The fitted curves look similar as with the untransformed data. There are few notable differences but generally the transformation has slightly improved the fits.      

\begin{figure}[!t]
\centering
\makebox{\includegraphics[width=0.80\textwidth]{f11}}
\caption{ Union-membership example. The $y$-axis represents the probability of union membership and the $x$-axis represents wages per hour. Five measurement error variance are considered; $\sigma_w^2=0$ (black), $\sigma_w^2=1$ (red), $\sigma_w^2=4$ (green), $\sigma_w^2=9$ (blue), and $\sigma_w^2= 16$ (purple). (a) The OSMEE with Gaussian sampling, (b) the OSMEE with deconvolution sampling , (c) Bayesian probit model, and (d) the OSMEE with Gaussian sampling on $\log$ wages.\label{fig:Union}}
\end{figure}

 \section{Extensions}\label{sec:extensions}
 
 There are many ways in which model
 \begin{align}
E(y_i|\gamma, s) &= \mu(\vect v_i^\top  \vect \gamma + f(x_i)),\tag{1}
\end{align}
can be extended. In each case, it is straightforward to estimate the mean and variance functions by simulation.  
For example, instead of modelilng only one variable nonparametrically, the nonparametric component can be the additive model
 \begin{align}
E(y_i|\gamma, s) &= \mu\left(\vect v_i^\top  \vect \gamma + \sum_{q=1}^Q f(x_{q,i})\right),
\end{align}
where some or all of $x_{1,i},\dots,x_{Q,i}$ are measured with error. 

In the case of longitudinal or multilevel data, suppose that $y_{i,j}$ is the $j$th measurement on the $i$ subject, and similarly for $\vect v_{i,j}$ and $x_{i,j}$.  Subject-specific effects can be modeled by replacing $\vect \gamma$ by
$ [\vect \gamma^\top \ \vect \gamma_j^\top]^\top$ where $\vect \gamma_j$ is a vector of subject-specific random effects and $\vect \gamma$ contains effects that are common to all subjects.

\section{Conclusions}
\label{sec:conc}
 In this article, we have proposed a methodology to adjust for measurement error in predictors when estimating regression functions nonparametrically. This method depends on modelling the mean and the variance of the response variable given the observed predictor and the random effects. Since, except for very limited cases, the exact mean and variance cannot be found we resorted to simulations to approximate these quantities before applying Taylor series expansion on the mean function to linearize the problem.  This method operates under various response distributions and link functions covering most if not all GLM family members including quasi-families as well. We have demonstrated through extensive simulation studies that this method works quite satisfactory under various scenarios and largely outperformed the Bayesian estimator.

\section*{Software}
The proposed methodology is implemented in the R package \texttt{gamME}, which is publicly available on CRAN and also provides broader functionality for measurement error modeling beyond the settings considered in this paper.

{}

\end{document}